\documentclass{aastex63}

\usepackage{wrapfig}

\submitjournal{ApJ}

\shorttitle{Silicon Technosignatures}
\shortauthors{Kopparapu et al.}
\graphicspath{{./}{figures/}}

\begin{document}

\title{ Detectability of Solar Panels as a Technosignature}

\correspondingauthor{Ravi Kopparapu}
\email{ravikumar.kopparapu@nasa.gov}

\author[0000-0002-5893-2471]{Ravi Kopparapu}
\affiliation{NASA Goddard Space Flight Center, 8800 Greenbelt Road, Greenbelt, MD 20771, USA}
\affiliation{SEEC, Sellers Exoplanet Environment Collaboration, NASA Goddard Space Flight Center, 8800 Greenbelt Road, Greenbelt, MD 20771, USA}
\affiliation{CHAMPs, Consortium on Habitability and Atmospheres of M-dwarf Planets}

\author[0000-0002-5060-1993]{Vincent Kofman}
\affiliation{NASA Goddard Space Flight Center, 8800 Greenbelt Road, Greenbelt, MD 20771, USA}
\affiliation{SEEC, Sellers Exoplanet Environment Collaboration, NASA Goddard Space Flight Center, 8800 Greenbelt Road, Greenbelt, MD 20771, USA}
\affiliation{American University, 4400 Massachusetts Avenue, NW, Washington, DC, 20016, USA}

\author[0000-0003-4346-2611]{Jacob Haqq-Misra}
\affiliation{Blue Marble Space Institute of Science, Seattle, WA, USA}
\affiliation{CHAMPs, Consortium on Habitability and Atmospheres of M-dwarf Planets}

\author[0000-0002-5893-2471]{Vivaswan Kopparapu}
\affiliation{Atholton High School, Columbia, MD 21044, USA}

\author[0000-0002-2685-9417]{Manasvi Lingam}
\affiliation{Department of Aerospace, Physics and Space Sciences, Florida Institute of Technology, Melbourne, FL 32901, USA}









\begin{abstract}
In this work, we assess the potential detectability of solar panels made of silicon on an Earth-like exoplanet as a potential technosignature. Silicon-based photovoltaic cells have high reflectance in the UV-VIS and in the near-IR,
within the wavelength range of a space-based flagship mission concept like the  Habitable Worlds Observatory (HWO). Assuming that only solar energy is used to provide the 2022 human energy needs with a land cover of $\sim 2.4\%$, and projecting the future energy demand assuming various growth-rate scenarios, we assess the detectability with an $8$ m HWO-like telescope. Assuming the most favorable viewing orientation, and focusing on the strong absorption edge in the ultraviolet-to-visible ($0.34-0.52$ $\mu$m), we find that several 100s of hours of observation time is needed to reach a SNR of 5 for an Earth-like planet around a Sun-like star at 10pc, even with a solar panel coverage of $\sim 23\%$ land coverage of a future Earth. We discuss the necessity of concepts like Kardeshev Type I/II civilizations and Dyson spheres, which would aim to harness vast amounts of energy. Even with much larger populations than today, the total energy use of human civilization would be orders of magnitude below the threshold for causing direct thermal heating or reaching the scale of a Kardashev Type I civilization. Any extraterrrestrial civilization that likewise achieves sustainable population levels may also find a limit on its need to expand, which suggests that a galaxy-spanning civilization as imagined in the Fermi paradox may not exist.

\end{abstract}


\section{Introduction} \label{sec:intro}

The search for extraterrestrial life has primarily focused on detecting biosignatures, which are remote observations of atmospheric or ground-based spectral features that indicate signs of life on an exoplanet. More recently, ``technosignatures'' referring to any observational manifestations of extraterrestrial technology that could be detected or inferred through astronomical searches has received increased attention \citep{Tarter07}. While the search for extra-terrestrial intelligence (SETI) through radio observations has been popular for decades, recent studies have proposed alternate searches for technosignatures in the UV to mid-infrared part of the spectrum: see \cite{NASA2018,LL2019,ML21,Socas-Navarro2021,jacob2022a} for a comprehensive description. 

 Specifically, methods to detect technosignatures through spectral signatures from exoplanets have been proposed as a means to utilize existing techniques and telescope facilities. These include nitrogen dioxide (NO$_{2}$) pollution \citep{kopparapu2021}, fluorinated compounds such as chloroflorocarbons (CFCs, \citealt{owen1980search,Lin2014,jacob2022b}), or nitrogen trifluoride (NF$_{3}$) and sulfur hexafluoride (SF$_{6}$) \citep{seager2023}, night-side city lights \citep{beatty2022} and agricultural signatures on exoplanets \citep{jacob2022c}. In this work, we focus on another potential technosignature: silicon solar panels. 

 Technological civilizations may harness their host star's radiation for their energy needs, just like our civilization has commenced with large scale solar photovoltaics. Most solar cells use silicon in different forms. \cite{ll2017} outlined three primary motivations for employing silicon-based solar panels, which might be broadly applicable. The first is the relatively high cosmic abundance of silicon compared to the elements utilized in other types of photovoltaics such as germanium, gallium, or arsenic. Second, the electronic structure of silicon (specifically its band gap) is well-suited for harnessing the radiation emitted by Sun-like stars \citep{SR16}. Third, silicon is also cost effective in terms of refining, processing and manufacturing solar cells \citep{bazilian2013}.\footnote{\url{https://www.energy.gov/eere/solar/solar-photovoltaic-cell-basics}}
 
 Based on these arguments, \cite{ll2017} suggest that the existence of large-scale silicon solar cells could produce artificial spectral ``edges'' in some UV wavelength bands when observing the atmosphere of an exoplanet in reflected spectroscopy because of the steep change in the reflectance of silicon. This artificial spectral edge may be similar to the vegetation red ``edge'' (VRE) seen between $0.70 - 0.75\,\mu$m that can be noticed in the reflected light spectrum of Earth \citep{sagan1993,arnold2002,woolf2002}. The ``edge'' refers to the noticeable increase in the reflectance of the material under consideration when a reflected light spectrum is taken of the planet. In the VRE case, the high reflectance arises due to the contrast between chlorophyll absorption at red wavelengths ($0.65 – 0.70\,\mu$m) and the scattering properties of the cellular and leaf structures at NIR wavelengths ($0.75-1.1\,\mu$m): see, for example, \cite{seager2005,turnbull2006,schwieterman2018,o2018} for more details. Detecting VRE on an exoplanet would provide contextual information about the type of widespread biological life (e.g., autotrophy), and corresponding atmospheric properties relevant for habitability. \cite{ll2017} suggest that a similar artificial edge, if manifested, could provide some contextual information about the kind of technological activity on a planet. 

Could we detect surface reflectance features of solar panels on exoplanets as technosignatures? While \cite{ll2017} suggested this possibility, they did not conduct any quantitative assessment of their detectablity. In this work,  we will consider the detectability of solar panels on an Earth-like planet around a Sun-like star with a LUVOIR-B (8 meter) class space telescope. The paper is structured as following: \S\ref{sec:methods} discusses the methods and models  used in this work. In \S\ref{sec:production}, we estimate the area of land that would need to be covered to provide human civilization with its energy needs today and in future scenarios. 
The potential detectability  is then assessed in \S\ref{sec:detectability}. A discussion section and a summary of our findings follows in \S\ref{sec:discussion}.

\section{Methods}
\label{sec:methods}
The methods to assess the detectability of photovoltaics as a signature for the presence of advanced civilizations are described here. In order to constrain the spectral signal, the following needs to be assessed:  1) The reflectivity of photovoltaics panels. 2) The panels need to be included in a suitable location on an Earth-ground map 
3) The spectroscopic signal from the panels need to be compared to simulations without the panels, and the signal-to-noise that can be attributed to the panels should be computed. 

\subsection{Silicon and the reflectivity as as spectral signature}\label{SSecSiRefl}
Pure silicon is not as well-suited (as a material) to be used in a photovoltaic cell since it is highly reflective in the ultraviolet-to-visible range. As electric energy is generated by absorbing a photon to promote an electron across the PN junction, any light that is reflected leads to a reduction in efficiency. To minimize reflection of light, photovoltaic cells are either subjected to texturing \citep{CG87,MCK04,KLK20} or coated in anti-reflective coatings, often TiO${_2}$ or Si$_3$N$_4$ \citep{ZG91,RGN11,SES20}; the coating results in the typical dark color seen on solar panels. 

In the scenario with the above anti-reflective coating, the artificial edge is still apparent, but less pronounced and deeper in the ultraviolet (with respect to pure silicon), when realistic materials are assumed for photovoltaic cells. For this work, the reflectance spectrum shown in Fig. \ref{fig:reflectance} is adopted. This explains a major source of divergence between \cite{ll2017} and our work, because the former emphasized greater surface coverage by pure silicon solar panels that could compensate for reduced efficiency, whereas we consider potentially more realistic photovoltaic cells endowed with higher efficiency, thereby warranting lower coverage.

\subsection{Generating the surface model containing solar panels}
Based on the estimated  surface area required for the current energy use discussed in \S\ref{sec:production}, a ground map is generated that hosts roughly 2.4\% of land coverage. The Sahara desert was fiducially chosen to host the solar panels. 
This region is both close to the equator, where a comparatively greater amount of solar energy would be available throughout the year, and has minimal cloud coverage.  However, dust storms are also prevalent, and have been increasing in frequency over the past four decades (see Fig.4, \citealt{varga2020}). Average events are $\sim 20$ per year with varying severity. Such events may reduce the available sunlight, further restricting the energy generated. We recognize and caution that no significant area of the Earth is uninhabited, and even the placement of solar farms in seemingly barren deserts has been contentious due to the destruction of the extremely fragile ecosystems that may consequently arise. However, our goal in this work is to assess the detectability of solar panels on an exoplanet, and as such, the most ``optimal'' land location in term of solar energy generation is chosen for this purpose. 

The ground surface model that is used for this study is based on the data products from the moderate resolution imaging spectroradiometer (MODIS), which is hosted on both the Terra and AQUA satellites operated by NASA Goddard Space Flight Center\footnote{\url{https://modis.gsfc.nasa.gov/about/}}. The MODIS-MCD12C1 maps provide yearly average coverage at high spatial resolution of 18 different land types across the entire planet. For this paper, ground coverage is reduced to 5 different types: ocean, snow/ice, grass, forest, and bare soil. The spatial resolution is binned down to 2.5 by 2 degree (longitude/latitude). The ground albedo, or the fraction of light that is reflected as a function of wavelength for the different surface types, are adopted from the United States Geological Survey database \citep{kokaly_usgs_2017}. Subsequently, solar panels are added as a sixth category, which are described with the reflectivity from RELAB (Reflectance Experiments Laboratory \citep{pieters_relab_2004}, and placed on the surface map at the chosen locations.

\subsection{Assessing the contribution to the reflectivity from silicon}
In order for a spectral feature to be detectable, it needs to satisfy two conditions. It has to be sufficiently strong, and its spectral signal needs to uniquely identifiable with the source molecule or in this case, ground coverage. The detectability of photovoltaic panels was proposed to be in the UV-VIS region of the spectrum, which is the range where the spectral feature of the panels are most uniquely identifiable. The infrared region is not suitable because the difference in the reflectivity here does not correspond to a spectrally unique feature: the features overlap with much stronger signals from the other surface components. This study is constructed so as to focus on the maximum detectability of the technosignature. This does not indicate that the signal may be fully uniquely attributed to the panels, which would require a follow-up study utilizing retrieval methods and an exhaustive search for possible overlapping signals (i.e. Rayleigh scattering, absorption by O$_3$ or hazes). The first step in assessing the detectability is finding the potential signal strength in the most optimistic case, which is pursued here.

\begin{figure} 
    \fig{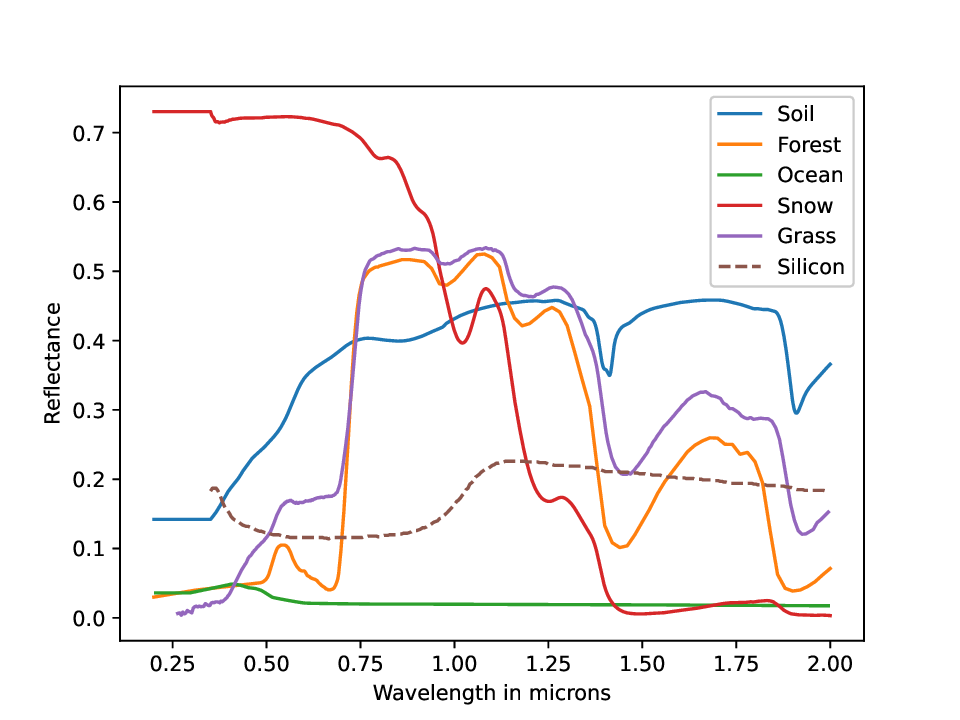}{0.55\textwidth}{}
        \caption{ Reflectance of different surface components on a planet as a function of wavelength. The reflectance of silicon-based photovoltaics is shown as dashed curve. The ground model is based on the data from MODIS dataset \url{https://modis.gsfc.nasa.gov/about/}. The solar cell reflectivity data is from RELAB (Reflectance Experiments
Laboratory \citep{pieters_relab_2004}. } 
    \label{fig:reflectance}
\end{figure}

It should also be noted that the placement of the panels in the desert is fortunate in terms of the detectability as well, because from the ground coverage components shown in Fig.\ref{fig:reflectance}, the contrast exhibited with soil is the second highest (after snow/ice).


\section{Photovoltaic Requirements for Earth}\label{sec:production}

At present, the power density (i.e., power generated per unit ground area) of solar energy is estimated to be $5.4$\,W\,m$^{-2}$\citep{millerkeith2018}.  Ignoring the effect on a habitable planet environment, if the total land area of Earth\footnote{\url{https://ourworldindata.org/land-use}} is 149 $\times 10^{6}$\,km$^{2}$, then the total power that {\it could be} generated if all the land were covered by solar panels would be $5.4\text{\,W\,m}^{-2}\times149 \times 10^{6}\text{\,km}^{2}= 804$ Tera Watts, or 25,374 exajoules per year. In 2022, the world power consumption from all primary energy sources (including commercially-traded fuels and modern renewables used to generate electricity) was 604 exajoules.\footnote{Page 8, ``Primary energy consumption'', \url{https://www.energyinst.org/statistical-review}} Clearly, all land need not be covered: only $\sim 2.4\%$ of land coverage by solar panels would be needed to match the world energy consumption in 2022.

\begin{figure}
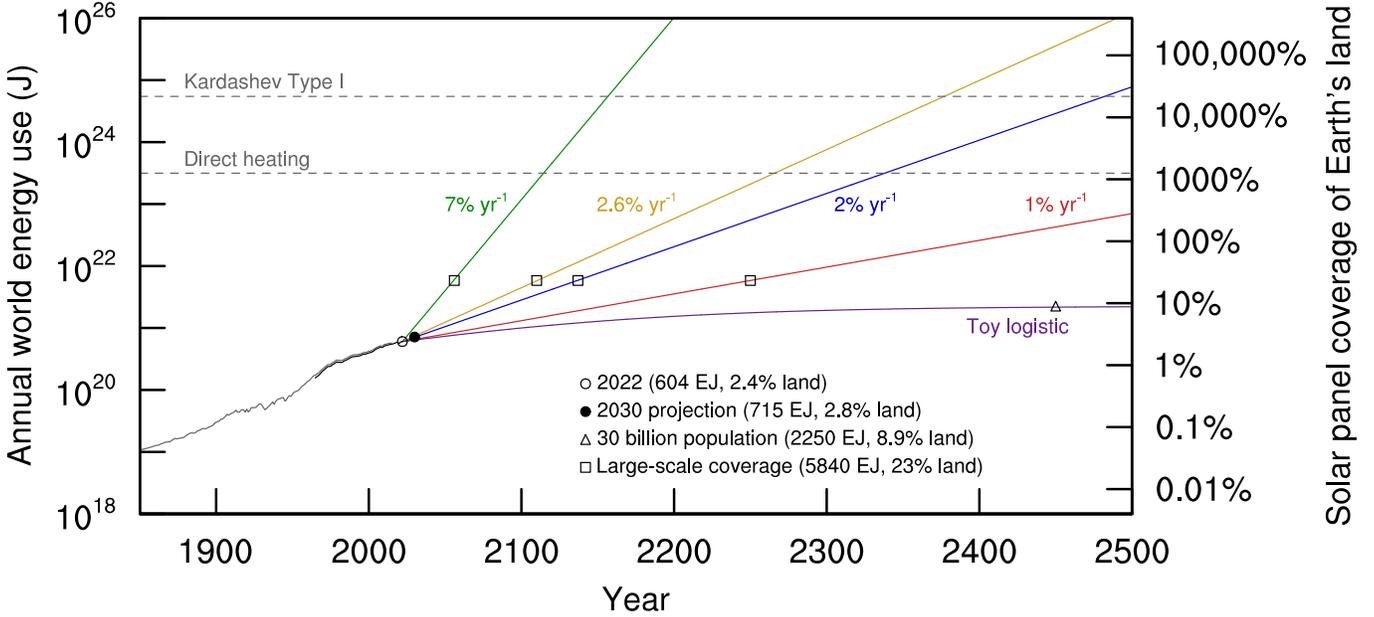
 
    \fig{fig-energyfuture-short.eps}{1.0\textwidth}{}
        \caption{Historic and projected annual world energy use in Joules (left axis) and in the corresponding fraction of Earth's land coverage by solar panels that would be required (right axis). Historic data are from the OECD (1850--2015, grey) and Energy Institute (1965-2022, black). The historic 2022 value is shown in an open circle and a projection for 2030 from the US Energy Information Administration is shown as a filled circle. Colored lines show future projections with constant growth at 7\%\,yr$^{-1}$ (green), 2.6\%\,yr$^{-1}$ (yellow), 2\%\,yr$^{-1}$ (blue), and 1\%\,yr$^{-1}$ (red). A hypothetical scenario of 23\% land coverage by solar panels is shown as open squares on each of these four future projections. An additional Toy logistic projection (violet) follows a 1\%\,yr$^{-1}$ growth rate up to the maximum energy use required to sustain a population of 30 billion people at a high standard of living, with the maximum value of the logistic function shown as an open triangle. Horizontal dashed grey lines indicate threshold values for causing direct heating of Earth's atmosphere ($\sim 10^{23}$ and reaching the Kardashev Type I limit (i.e. harvesting all available solar energy). Calculations assume a fixed $5.4$\,W\,m$^{-2}$ solar panel power density.} 
    \label{fig:energy}
\end{figure}

Figure \ref{fig:energy} shows historic annual world energy use in Joules, and projected energy usage under various growth-rate scenarios. This plot is similar to the growth rate figure shown in \cite{mullan2019}, where the authors discussed the implications of population growth as related to the energy usage. Two data sets are shown: one from the Organization for Economic Cooperation and Development (OECD) data\footnote{\url{https://www.oecd-ilibrary.org/energy/primary-energy-supply/indicator/english_1b33c15a-en}} from 1850--2015, and another curve from the Energy Institute\footnote{Page 8, ``Primary energy consumption'' \url{https://www.energyinst.org/statistical-review}} from 1965--2022. The purpose of representing both datasets is to show that the two datasets generally agree, apart from a small systematic difference. Projected energy usage in 2030 is based on estimates from the US Energy Information Administration\footnote{\url{https://www.eia.gov/outlooks/archive/ieo09/world.html}} (678$\times 10^{15}$\,Btu $\approx$ 715 exajoules), and the corresponding land coverage needed ($2.8\%$) to power the world entirely on solar panels is also shown. 

Future projections are shown in Figure \ref{fig:energy} by assuming scenarios of constant growth of 7\%\,yr$^{-1}$, 2.6\%\,yr$^{-1}$, 2\%\,yr$^{-1}$, and 1\%\,yr$^{-1}$ based on the 2022 world energy consumption estimate of 604 exajoules from the Energy Institute, and assuming a fixed solar power density value of 5.4 W m$^{-2}$. The historical period from 1850--2022 shows an average growth rate of about 2.6\%\,yr$^{-1}$, although this growth rate decreased to about 2\%\,yr$^{-1}$ during the more recent 1965--2022 period. A growth rate of 7\%\,yr$^{-1}$ was used by \citet{von1975population} in a previous analysis of limits to growth; such a projection was consistent with the more rapid growth rate observed from the period of about 1950--1975. These scenarios, along with a more conservative 1\%\,yr$^{-1}$ scenario and a toy logistic scenario, serve to illustrate the range and inherent uncertainties in making such projections about the future. 

Figure \ref{fig:energy} also indicates the point on each projection where the energy demands would require a solar panel coverage of 23\% of Earth's land---about the size of Africa---with an annual energy use of 5840 exajoules. This magnitude of energy use would occur by 2056 for the 7\%\,yr$^{-1}$ scenario, by 2110 for the 2.6\%\,yr$^{-1}$ scenario, by 2137 for the 2\%\,yr$^{-1}$ scenario, and by 2250 for the 1\%\,yr$^{-1}$ scenario. The idea of covering such a large fraction of Earth's surface with solar panels would inevitably have many undesirable consequences on climate and local environments; nevertheless, this 23\% land coverage limit will be used as an upper limit in the detectability calculations in \S\ref{sec:detectability}.

It is also worth noting that the energy output from 23\% land coverage would far surpass that required to provide all people with a high standard of living. The analysis by \cite{jackson2022human} found that human well-being may peak at a per capita energy use of 75 gigajoules per person,\footnote{Many other studies have also concluded that an annual energy consumption of $\lesssim 100$ GJ per person should suffice to achieve a high standard of living \citep{ACL16,MSR20,VSO21}.} so the 5840 exajoules generated by 23\% land coverage would be more than adequate to sustain a population of 10 billion people (corresponding to 750 exajoules, or 3\% land coverage)---which is approximately the maximum human population predicted by most United Nations models. For the purpose of illustration in this study, the energy use of a human population of 30 billion people (2250 exajoules, 8.9\% land coverage) is used as the upper bound of the toy logistic function in Figure \ref{fig:energy}. This toy logistic function serves to demonstrate the possibility of a nonlinear trajectory in future energy use, which may even reach a stable equilibrium. For human civilization today, it is encouraging to consider the possibility that 8.9\% solar panel coverage of Earth's land---approximately the size of China and India combined---would be more than enough to provide a high quality of life in the future. The implications of this possibility for technosignatures will be discussed in \S\ref{sec:discussion}. 





\section{Detectability Requirements for Photovoltaics}\label{sec:detectability}

\begin{figure}
   \gridline{\fig{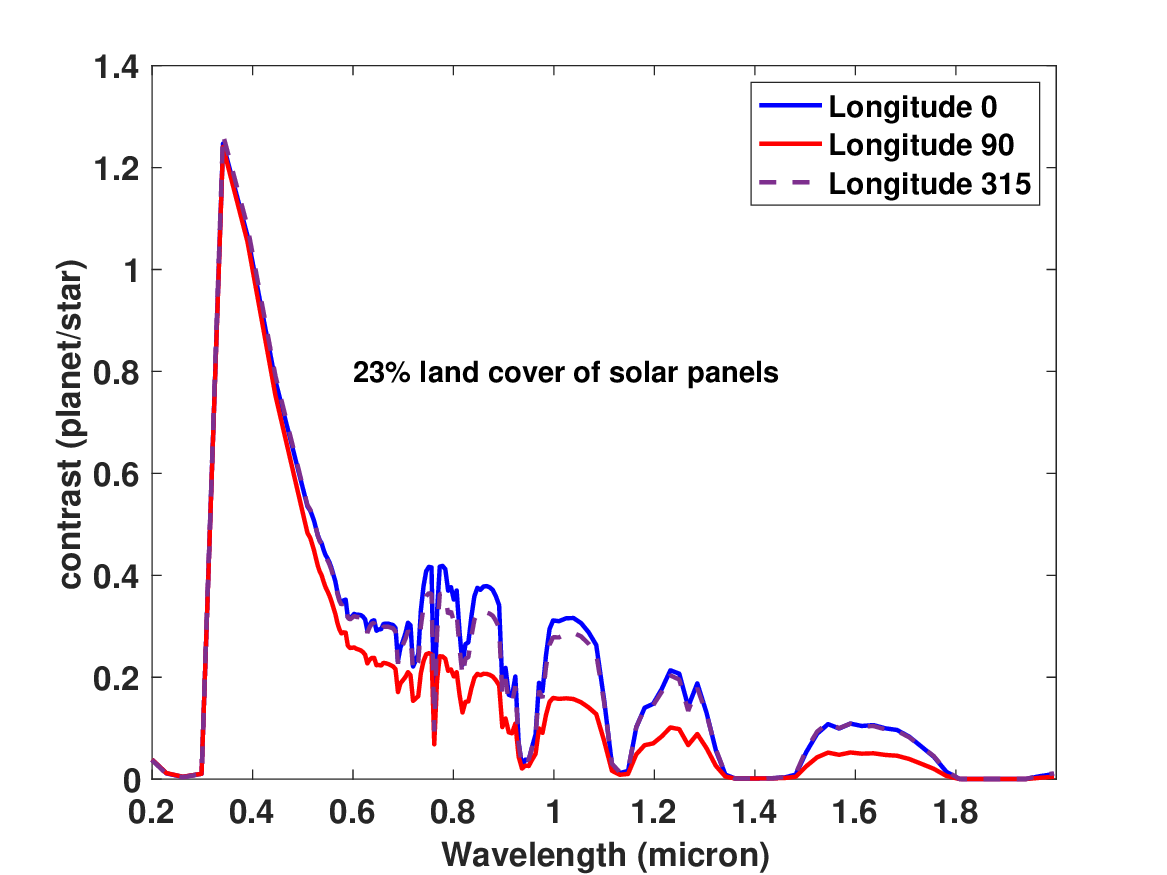}{0.49\textwidth}{(a)}
          \fig{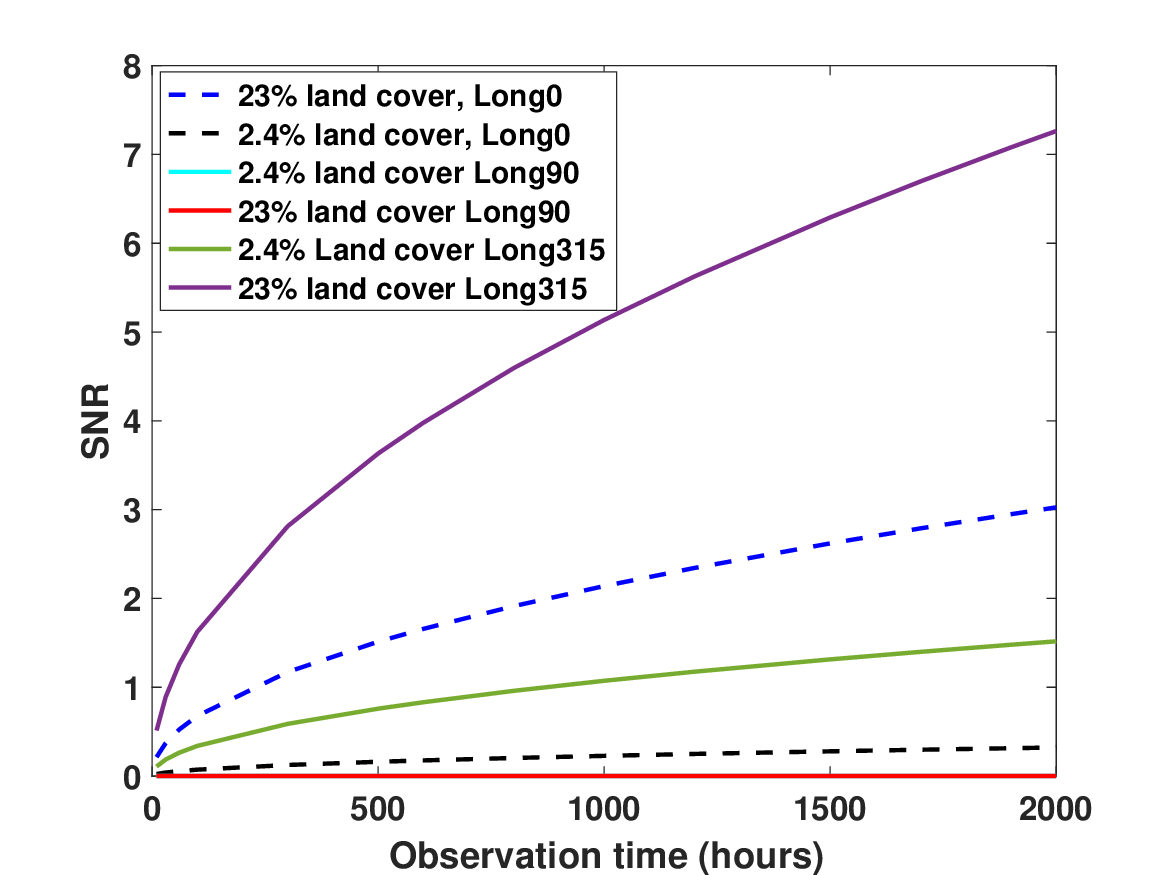}{0.49\textwidth}{(b)}
          }
    \caption{(a) Reflectance spectra (expressed in terms of the planet to star contrast ratio) for different planetary orientations with respect to the observer for a $23\%$ land cover of solar panels. We choose the $0.34 - 0.52\mu$m range to estimate the detectability of Si-based solar panels, as discussed in \S\ref{sec:detectability} and shown in Fig.\ref{fig:reflectance}.  Within this range, there is a noticeable difference in the spectra, which contributes to higher SNR for an observing longitude of $315^{\circ}$ (fully visible panel coverage). (b) SNR of detecting silicon solar panels as a function of observation time with a 8m class LUVOIR-B-like telescope for different orientations of the planet towards the observer. A longitudinal orientation of $315^{\circ}$ yields a higher SNR because the solar panels on the planet are better oriented towards the observer, as compared to a longitude of $90^{\circ}$ (partially visible).} 
    \label{fig:snr}
\end{figure}

We performed simulations with the Planetary Spectrum Generator (PSG: \citealt{psg:2018,villanueva_fundamentals_2022}) to generate reflected light spectra, and then calculated the required signal-to-noise ratio (SNR) to detect the signature of solar panels in the $0.34\,\mu$m--$0.52\,\mu$m range, which overlaps with the UV and optical spectral region for HWO. We selected this particular region because of the following reason: Fig. \ref{fig:reflectance} shows that the reflectance of Silicon solar panel has peaks below $\sim0.4\,\mu$m and between 1-2$\,\mu$m. However, these signatures  strongly overlap with features from other surface materials. Any change in the coverage of solar panels will be obscured by changes in the relative amount of ocean versus soil or vegetation coverage, as can be see in Fig. \ref{fig:snr}.  Fig. \ref{fig:snr}a shows the spectral contrast ratio versus wavelength for different viewing angles from the point of view of an observer.  The contrast ratio is greater than 1 because we are presuming a coronagraph which blocks most of the starlight. The changes in the spectra beyond 0.8$\,\mu$m are predominantly changes in the reflectivity of the ground coverage. Any contribution from the silicon is blended into the overall spectra in this wavelength region; note that the overlap of the silicon features with molecular absorption features is not the focus of our investigation. Our analysis provides an upper limit to the signal that may originate from photovoltaic panels. The potential overlap with molecular features would be the next step in the investigation in case the signal would be detectable.

Fig. \ref{fig:snr}b shows the SNR values of different land coverage of solar panels as a function of observation time for a 8m HWO-like telescope. The model instrument set is similar to the LUVOIR concept study telescope, as the design of HWO mission is currently ongoing. As both LUVOIR-B and the HWO are concepts for off-axis telescopes, the LUVOIR-B concept is the current best approximation. It has an internal coronagraph with the key goal of direct exoplanet
observations \citep{JZP22}. It is equipped with three channels: NUV (0.2–0.525 $\mu$m), visible (0.515–1.030 $\mu$m) and NIR (1.0–2.0 $\mu$m).  The NUV channel is capable of high-contrast imaging only, with an effective spectral resolution of R$\sim 7$. The optical channel contains an imaging camera and integral field spectrograph (IFS) with R=140. The planet is kept at a quadrature phase (in our simulation, an orbital phase angle of $270^{\circ}$). This is an Earth-like planet at an Earth-like distance around a Sun-like star at distance of $10$ pc. 

The wavelength dependent SNR is calculated as the difference between the spectra with and without the solar panels divided by the noise simulated by PSG for the instrument under consideration (see section 5.3 of \citet{psg:2018}, chapter 8 in \citet{villanueva_fundamentals_2022}, and also the PSG website,\footnote{\url{https://psg.gsfc.nasa.gov/helpmodel.php}} where the noise model is discussed in detail).  The ``net SNR'' is calculated by summing the squares of the individual SNRs at each wavelength within a given band (either NUV or VIS), and then taking the square root. This methodology  is largely robust to SNR, as long as the feature is resolved by the spectrum. The SNR is calculated at different longitudinal views of an Earth-like planet from observer's view point. Only three longitudinal views are shown for illustrative purpose. A $0^{\circ}$ longitude represents a viewing angle where the placement of the solar cells are only visible partially from the observer's view point. A 315$^{\circ}$ longitudinal view indicates the placement of solar cells in almost the full view of the observer. A $90^{\circ}$ longitude represents a viewing angle where solar cells are not visible to the observer. 

\begin{figure}
    \fig{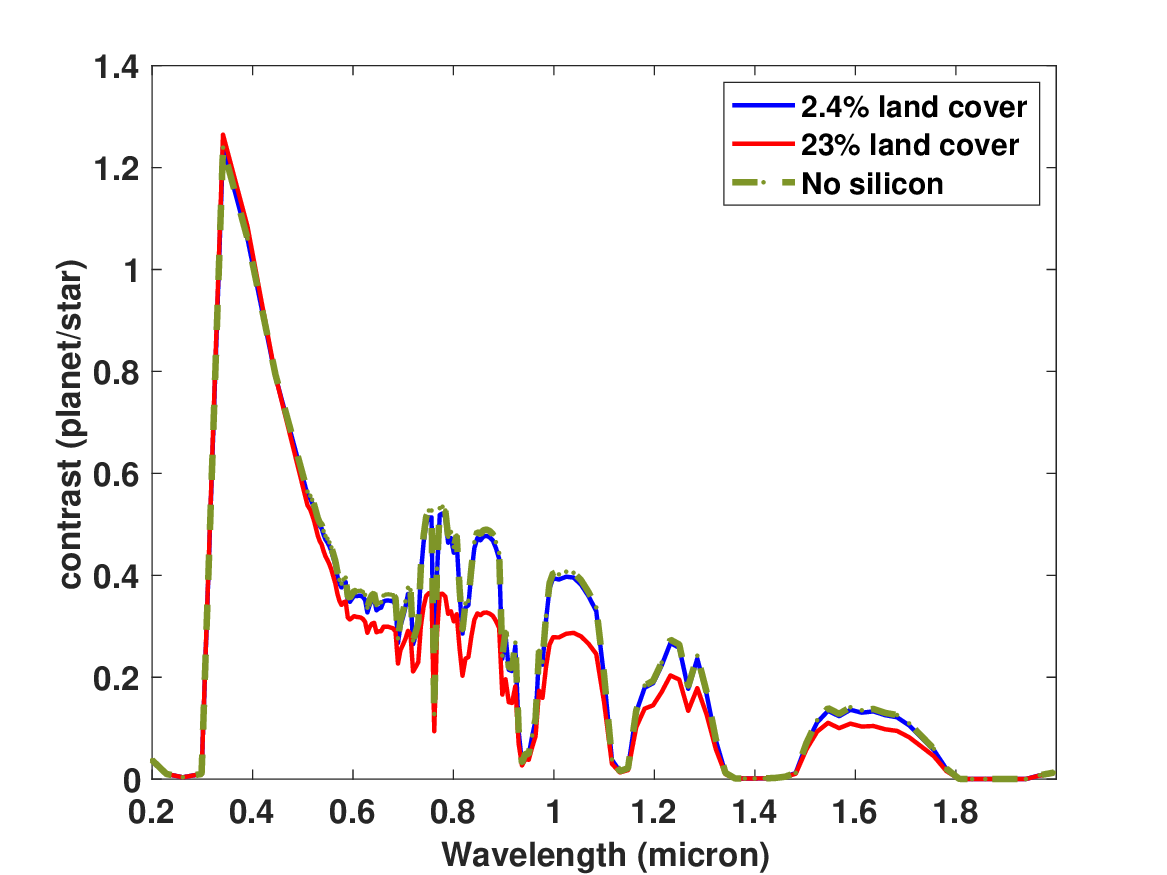}{0.55\textwidth}
    {}
    \caption{Planet-star contrast ratio as a function of wavelength for $2.4\%$ (blue solid), $23\%$ (red solid) and $0\%$ (green dashed) land coverage of solar panels. The planet longitudinal orientation is $315^{\circ}$, where the solar panels are oriented towards the observer. Only the $23\%$ land cover displays any significant variation in the spectrum. This trend is also manifested in the SNR curve of Fig.\ref{fig:snr}, where the maximum SNR is obtained for a $23\%$ land cover oriented at $315^{\circ}$ planet longitude.} 
   \label{fig:s2}
\end{figure}

 Fig. \ref{fig:snr}b indicates that even with the most ambitious land coverage ($\approx 23\%$), and with a favorable viewing perspective to the observer (planet longitude of $315^\circ$), it would take several hundred hours of observation time in reflected light spectra with an 8\,m size telescope to reach SNR $\sim 5$ (solid purple curve). This result can be understood from inspecting Fig. \ref{fig:s2}, where the spectra in planet-star contrast ratio are plotted for three cases: $2.4\%$ (blue solid), $23\%$ (red solid) and zero (dashed green) land coverage of silicon panels. Because we have chosen the $0.34\,\mu$m--$0.52\,\mu$m range to calculate the impact of silicon panels on the reflectance spectra, the difference between a planet with and without silicon is not markedly different, even with $23\%$ land cover, as can be seen from Fig. \ref{fig:s2}. This suggests that the artificial silicon edge suggested by \citet{ll2017} may not be detectable. The discrepancy partly stems from the choice of photovoltaic cell properties, as indicated in Section \ref{SSecSiRefl}, because the reflectance of realistic solar cells is less pronounced than pure silicon, the latter of which was evaluated in \citet{ll2017}.  A larger coverage than $23\%$ would result in slightly lower observation times; however, refer to Section \ref{sec:discussion} for the implications of larger coverage by photovoltaic cells.
 

\section{Discussion} \label{sec:discussion}
The results from the previous section seem to indicate that even ambitious deployment of solar panels (constructed with silicon and hosting anti-reflective coatings) covering a significant land area of an Earth-like exoplanet may not be enough to be detectable with a 8\,m HWO-like telescope. These calculations presumed a fixed-present day efficiency for solar panels, so any technological improvements in efficiency would only decrease the required land coverage, and consequently decrease detectability. 

\citet{ll2017} conjectured that tidally locked planets around M-dwarfs might be a suitable place to start looking for photovoltaic cells. Unfortunately, due to the close proximity of planets in the habitable zone around M-dwarfs -- which poses issue for spatially resolving the planet -- currently there is no technological pathway to directly observing these types of planets. In the case that this setup would be possible though, the detectability of the silicon UV edge on planets around M-dwarfs would be further diminished, since the stellar emission in the relevant range  ($0.34-0.52\,\mu$m) is significantly lower. 


Note that we have performed all of the SNR calculations based on an orbital phase angle of $270^{\circ}$ with edge-on orbital configuration (i.e., inclination $=90^{\circ}$). Additional calculations varying the phase angle, varying the inclination angle, and so forth are needed to fully assess the potential detectability. Furthermore, we have not explored the extent of the effect of varying mirror size on the SNR, or a different stellar host spectral type (like a K-dwarf star).  \cite{bk2019} generated composite maps of orbiting photovoltaic power sources if they were situated around a planet like Proxima Centauri b. Specific simulations and the corresponding SNR estimates relevant for HWO from such orbiting structures needs to be undertaken. These additional considerations are left for future work.

As Earth remains the only example of an inhabited planet with a technosphere, future trajectories of Earth can provide insight into the constraints that might also apply to extraterrestrial technospheres. It is worth reiterating that only 3\% land coverage by solar panels would be needed to support a population of $10$ billion people at a high standard of living, and the toy logistic curve shown in Figure \ref{fig:energy} indicates that even a population of $30$ billion people at a high standard of living would require far less energy than the power output of the 23\% land coverage scenario used in these detectability calculations. These estimates not only do not account for increased efficiency due to technological innovations, but they also do not account for other sources of energy. Hence, the actual solar requirements for Earth would likely be much reduced, especially in case controlled nuclear fusion becomes viable. For Earth, these results suggest the possibility of a future in which the energy needs of even a larger-than-predicted population could be fully met with known technology.  

If Earth is taken as an example in the search for technosignatures, then this raises the question: Is the large-scale deployment of solar panels on a planetary surface ever needed? Any actual large-scale deployment would certainly raise numerous issues regarding the feasibility of such a deployment or the logistical challenges in global distribution of energy, but a more fundamental unknown is whether a technological civilization would ever require such large energy demands to justify a scenario such as 23\% land coverage or greater. Several of the projections shown in Figure \ref{fig:energy} reach a time at which dissipation from energy use exceeds the threshold for contributing significant direct heating to Earth's climate ($\sim 3\times 10^{23}$\,J), which occurs in the year 2265 for the $2.6\%$ yr$^{-1}$ growth rate scenario and 2338 for the $2\%$ yr$^{-1}$ growth rate scenario. Yet, it is not evident that such vast energy requirements will ever be necessary on Earth: the energy requirements for the global human population to afford a high standard of living fall several orders of magnitude below the direct heating threshold. The motivation to prevent direct heating of the atmosphere may serve as an additional deterrent to avoid such scenarios, and further suggests that only modest-scale deployment of solar panels would ever be needed to achieve sustainable development goals on Earth. 

Speculating even further shows that these scenario projections will reach the limit of a Kardashev Type I civilization \citep{1964SvA.....8..217K}, which is a civilization that uses all available energy on its planet (i.e., all starlight incident at the top of the atmosphere). This Type I threshold ($\sim 5\times 10^{24}$\,J) occurs in the year 2377 for the $2.6\%$ yr$^{-1}$ growth rate scenario and 2482 for the $2\%$ yr$^{-1}$ growth rate scenario. The equivalent land area required to meet such energy demands with solar power exceeds 100\% in these cases, which indicate that other sources of power---possibly including space-based solar power---would be required in these scenarios. Such speculative scenarios suggest images of ``Dyson spheres/swarms'' \citep[e.g.,][]{dyson1960search,wright2020dyson} of solar collectors that expand a civilization's ability to capture the energy output of its host star. 

\citet{1964SvA.....8..217K} even imagined a Type II civilization as one that utilizes the entirety of its host star's output; however, such speculations are based primarily on the assumption of a fixed growth rate in world energy use. But such vast energy reserves would be unnecessary even under cases of substantial population growth, especially if fusion and other renewable sources are available to supplement solar energy. The concept of a Type I or Type II civilization then becomes an exercise in imagining the possible uses that a civilization would have for such vast energy reserves. Even activities such as large-scale physics experiments and (relativistic) interstellar space travel \citep[cf.][Chapter 10]{ML21} might not be enough to explain the need for a civilization to harness a significant fraction of its entire planetary or stellar output. In contrast, if human civilization can meet its own energy demands with only a modest deployment of solar panels, then this expectation might also suggest that concepts like Dyson spheres would be rendered unnecessary in other technospheres.

These results also have implications for the problem known as the Fermi paradox \citep{webb2015if,cirkovic2018great,forgan2019solving} or Great Silence \citep{brin1983great}: if extraterrestrial expansion through the galaxy is relatively easy, then where are they? In the context of limits to growth, the recognition that human civilization could reach a sustainable equilibrium in its population and energy use that falls well below the Type I threshold suggests that extraterrestrial civilizations may not be compelled to expand for reasons of subsistence. This conclusion mirrors the ``sustainability solution'' to the Fermi paradox \citep[e.g.,][]{haqq2009sustainability,mullan2019}, which suggests that any extant extraterrestrial civilizations only expand proportionally with their planet's carrying capacity. Any civilization that is able to achieve sustainable population levels with high standard of living may also settle on a limit on any need to expand, which may likewise constrain the magnitude of any potential technosignatures generated by such a civilization. 

Underlying this discussion are two competing philosophical positions regarding the tendency of life to expand. The first position assumes that (a) life will evolve to utilize the maximum energy available in its environment, while the second assumes that (b) life will evolve to utilize as much energy as needed in its environment to reach an optimal level of existence. To the extent that the environment will always impose thermodynamic limits, the second option (b) may probably be favored by life. But does life tend to expand up to such thermodynamic limits if unopposed, or will life tend to optimize its consumption of resources rather than maximize them? Such interdisciplinary questions highlight the intersection of technosignature science with concepts from ecology \citep[e.g.,][]{meurer2024astroecology}, and further scrutiny of the proposed general tendency of life to expand may be useful for thinking about and enivisioning the particular tendencies for technospheres to expand.

\section{Conclusion} 
\label{sec:conclusion}
We analyzed the detectability of silicon-based solar panels (with anti-reflective coatings) as a signature of extraterrestrial technology. Assuming a 8-meter HWO-like telescope, focusing on the reflection edge in the UV-VIS, and considering varying land coverage of solar panels on an Earth-like exoplanet that match the present and projected energy needs, we estimate that several hundreds of hours of observation time is needed to reach a SNR of $\sim 5$ for a high land coverage of $\sim 23\%$. 

We subsequently assessed the need for the deployment of such large-scale solar panels, taking into consideration various projected growth rates of future energy consumption. We find that, even with significant population growth, the energy needs of human civilization would be several orders of magnitude below the energy threshold for a Kardashev Type I civilization, or a Dyson sphere/swarm which harnesses the energy of a star. This line of inquiry reexamines the utility of such concepts, and potentially addresses one crucial aspect of the Fermi paradox: we have not discovered any large scale engineering yet, conceivably because advanced technologies may not need them.

\acknowledgments
J.H.M., M.L., and R.K.K. gratefully acknowledge support from the NASA Exobiology program under grant 80NSSC22K1009. The authors also acknowledge support from the Goddard Space Flight Center (GSFC) Sellers Exoplanet Environments Collaboration (SEEC), which is supported by the NASA Planetary Science Division's Research Program. Any opinions, findings, and conclusions or recommendations expressed in this material are those of the authors and do not necessarily reflect the views of their employers or NASA.

%

\vspace{5mm}







\bibliography{sample63}{}
\bibliographystyle{aasjournal}



\end{document}